# An Ensemble Framework For Day-Ahead Forecast of PV Output Power in Smart Grids

*Muhammad Qamar Raza, Student Member, IEEE, N. Mithulananthan, Senior Member, IEEE, Jiaming Li, Kwang Y. Lee, Fellow,IEEE, Hoay Beng Gooi, Senior Member, IEEE*

***Abstract*-- The uncertainty associated with solar photo-voltaic (PV) power output is a big challenge to design, manage and implement effective demand response and management strategies. Therefore, an accurate PV power output forecast is an utmost importance to allow seamless integration and a higher level of penetration. In this research, a neural network ensemble (NNE) scheme is proposed, which is based on particle swarm optimization (PSO) trained feedforward neural network (FNN). Five different FFN structures with varying network complexities are used to achieve the diverse and accurate forecast results. These results are combined using trim aggregation after removing the upper and lower forecast error extremes. Correlated variables namely wavelet transformed historical power output of PV, solar irradiance, wind speed, temperature and humidity are applied as inputs to the multivariate NNE. Clearness index is used to classify days into clear, cloudy and partial cloudy days. Test case studies are designed to predict the solar output for these days selected from all seasons. The performance of the proposed framework is analyzed by applying training data set of different resolution, length and quality from seven solar PV sites of the University of Queensland, Australia. The forecast results demonstrate that the proposed framework improves the forecast accuracy significantly in comparison with individual and benchmark models.**

***Acronyms***

| | |
|---|---|
| AI | Artificial Intelligence |
| ANN | Artificial Neural Network |
| AR | Auto-Regressive |
| ARIMA | Autoregressive Integrated Moving Average |
| BPNN | Back-Propagation Neural Network |
| CorC | Correlation Coefficient |
| CPU | Central Processing Unit |
| EN | Ensemble Network |
| ENF | Ensemble Framework |
| ENP | Ensemble Predictors |
| FNN | Feedforward Neural Network |
| LM | Levenberg–Marquardt |
| MAE | Mean Absolute Error |
| MAPE | Mean Absolute Percent Error |
| NN | Neural Network |
| PSO | Particle Swarm Optimization |
| PV | Photovoltaic |
| SG | Smart Grid |
| WT | Wavelet Transform |

Muhammad Qamar Raza and N. Mithulananthan are with the Power and Energy System group, at School of Information Technology and Electrical Engineering, University of Queensland, Brisbane, QLD 4072, Australia (e-mail: qamar.raza@uq.edu.au; mithulan@itee.uq.edu.au).
K. Y. Lee is with Department of Electrical & Computer Engineering, Baylor University, Waco, TX 76706, USA (e-mail: Kwang_Y_Lee@baylor.edu).
Jiaming Li is associated with Data61, Commonwealth Scientific and Industrial Research Organization (CSIRO), Sydney, Australia (jiaming.li@csiro.au).
H. B. Gooi is with the Department of EEE, Nanyang Technological University, Singapore 639798 (e-mail: ehbgooi@ntu.edu.sg).

## I. Introduction

The installed capacity of solar photo-voltaic (PV) power has been increasing steadily and growing in several countries in multiple straight years. The PV installations at different scales produce large economic benefits to consumers and market operators. The PV power output is not steady due to the influence of various factors. Two crucial factors affecting the power output are solar radiations on PV panel and air temperature. In addition, other factors which influence the PV power output are speed and direction of the wind, cloud cover, relative humidity, hourly solar angle, and location and orientation of solar PV arrays [1]. However, uncertain and variable solar power output introduces different direct and indirect challenges to power systems and tends to reduce economic benefits. Hence, to address the challenges associated with increased penetration of solar PV, there is a necessity to design an accurate forecast model for PV power output over a range of forecast horizon.

There is a range of methods available for PV power output prediction. These methods can be classified as physical, persistence, statistical and combined approaches. Most of the forecast methods utilize the historical data, meteorological and other exogenous variables as the input to the multivariate forecast model. A number of techniques are proposed to predict the solar PV power output resulting from large penetration and its potential in meeting the shortage of energy [2-4]. In [5] and [6], authors proposed a time-series PV output forecast model for higher forecast accuracy. In [4], different learning techniques such as adaptive neuro-fuzzy, auto-regressive integrated moving average (ARIMA) and *k*-nearest neighbor (KNN) search were used in developing a numerical weather prediction (NWP) model. In order to examine the performance of the model, a variable forecast horizon from 1 to 39 hours was considered. In [7] and [8], a weighted quantile regression, Markov chain and vector autoregression framework methods were used for probabilistic forecasting. In [9], a combined technique for one-hour ahead forecast based on three models named as a self-organizing map, learning vector quantization and fuzzy inference system was presented. In another research, the authors proposed a combined model to predict the global solar radiation using Extreme Learning Machine algorithm and Coral Reefs Optimization with different meteorological variables [10]. Nevertheless, there is still an opportunity to enhance the forecast accuracy, which could be used for a number of other applications. In addition, there is a need to investigate the impact of the length and quality of training data on forecasting performance.

In [11], the authors designed a forecast framework based on three different mathematical models and compared it with a neural network (NN) model. According to the authors' findings, the NN model outperformed the conventional mathematical models with better forecast accuracy. However, it becomes necessary to further investigate the impact of different NN forecast models and the associated number of inputs on forecast output. In [12], the authors designed a model that combines the performances of artificial neural networks (ANN) and the regression model. The proposed ANN model demonstrated a slightly higher forecast accuracy than the regression model with temperature and solar radiation as input variables. In another research [13], authors compared performances of the PV output forecast of five different models namely the ANNs, ARIMA, KNN, ANN trained with the hybrid Genetic algorithm, and a persistent model for 1 MW PV plant in California. The ANN prediction model resulted in less forecast error, up to 11.42 root-mean square error (RMSE) for 1-h ahead forecasting. However, the NN forecasting performance is largely affected by the network training and noisy model inputs. Generally, gradient-based learning techniques such as backpropagation-based techniques are used to train NN. In [14], the authors designed a NN model and trained it with backpropagation (BP) learning technique. In the NN model, historical PV output and meteorological variables along with an aerosol index (AI) were used to improve the forecast accuracy. The BP-trained NN model does not provide a better forecast accuracy. There is a possibility that the network can be trapped in local minima in the learning process with the BP. This leads to poor network training, affecting the forecast output as a result. The prediction accuracy of a model can be further enhanced by optimizing the NN learning and overcome the shortcomings of the stand-alone NN.

It can be possible by exploring a number of possible ways to accommodate the effects on solar PV output and by enhancing the capability of NN predictors. In previously reported research, a combined model was implemented to accurately forecast with two or more models that depend on each other. However, the overall forecasting accuracy of the model was affected due to the bad performance of an individual model within the hybrid framework. Therefore, there is a room to redesign the model, in which each predictor is independent of each other's performance. In [15], authors designed a forecast framework based on multi-predictors, which are organized in a systematic manner to combine the output of all predictors.

The variable performance of individual predictors produce different forecast results for the same training data and input variables. The varied outputs of the independent predictors individually improve the overall accuracy by exploring different possible solutions. The neural network ensemble (NNE) is a method to combine the output of multiple models and possibly increase the forecast accuracy by not relying on a single model. In the NNE method, multiple NN's are created and organized in a systematic manner. The output of all individual NN is combined to produce an output. Such a combination of networks is popularly referred to as fusion or aggregation. The NNE depicts improved prediction results in load demand forecast application in our earlier study [16].

### A. *Contributions:*

In this paper, the NNE framework based on the feedforward neural network (FNN) is proposed. The FNN is selected with a number of performance case studies in comparison with the Back-Propagation Neural Networks (BPNN), Radial Basis Function (RBF) network, and Elman Network (ENN). The wavelet transformation (WT) is applied to historical PV power output data to remove the spikes and peaks. Historical PV output data with meteorological variables are applied to the forecast model as inputs. Each FNN is trained using PSO technique in the NNE. Each FNN generates the forecast output in the ensemble network (EN) and the output of each network is combined using a trim aggregation technique. In addition, it can be further improved by carefully applying optimal training data with different resolution, length and quality. The major contributions of the proposed paper can be highlighted as follows:

1) Design and integration of the neural predictors with varying structures in a forecasting framework along with the WT inputs.
2) Aggregation of the EN predictors using a trim technique for better forecast output by removing forecast error extremes.
3) Training of neural predictors using the PSO for better network training to enhance the forecast accuracy of the individual NN model.
4) Incorporation of historical PV power output data transformed with Wavelet with meteorological variables such as humidity (H), temperature (T), wind speed (WS) and solar irradiance as proposed forecast framework inputs.
5) Classification of the days into the clear, cloudy and partial cloudy days based on clearness index to analyze the performance of the NNE in each season.
6) Investigate the performance of the proposed framework under different length and quality of training data.

The rest of paper is organized as follows: Section II describes characteristics of the PV power output profile. Section III presents the data preparation with WT and FNN. The methodology of the proposed framework, trim aggregation technique, and benchmark model are explained in Section IV. Results and discussion are presented in Section V and conclusions of the paper are summarized in Section VI.

## II. Characteristics of PV Power Output

Solar irradiance and temperature largely affect PV power output. The power output (PO) of a single PV array at standard test conditions (STC) at the maximum power point tracking (MPPT) condition is $P_{O,STC}$ and can be calculated as Eqs. 1 and 2 [9]:

$$P_{O,STC} = \left[ P_{PV,STC} \frac{SI}{1000} \left[ 1 - \gamma \left( T_x - 25 \right) \right] \right] N_{P,P} N_{P,S} \tag{1}$$

$$T_x = T_{amb} \frac{SI}{800} \left( N_{OCT} - 20 \right) \tag{2}$$

where $T_{amb}$, $N_{P,S}$, $N_{P,P}$, $SI$, $T_x$, $N_{OCT}$, and $P_{PV,STC}$ are ambient temperature, number of PV arrays in series, number of PV arrays in parallel, solar irradiance, temperature at maximum power point (MPP), a constant term, and the PV power output at MPP and STC.

## *B. Solar PV Sites and Data Preparation*

A total of 1.22MW capacity of solar PV is installed in different buildings at the University of Queensland (UQ), Australia. In this research, the solar and meteorological data sets at UQ St. Lucia and Gatton campuses are utilized to train and test the proposed scheme [17]. The aggregated power of Sir Lew Global Change Institute (GCI), Edwards Building (LEB), Advanced Engineering Building (AEB), Car Park 1 (CP1), UQ Centre (UQC), Car Park 2 (CP2) and Gatton Solar Research Facility Dual Axis (GSRFDA) is used [17]. One-year data from January 1 to December 31, 2014, from UQ Solar data management system (UQ-SDMS), is used. The UQ-SDMS has real time recorded data of solar output from 7 am to 5 pm. The PV power output for other than 7 am to 5 pm is considered as zero, as these values are not significant due to unavailability of sunlight. The prepared 1-min, 5-min and 30-min datasets contain 7300, 43,800 and 219,000 measurements, respectively. However, 0.62% of data is missing in the real-time recorded data, which could be resulted from a number of reasons such as error in the sensor (instrumental) and data management system. Therefore, the missing data is reconstructed using the similarity technique. In addition, 1, 15, 30 and 60 min datasets are prepared to train the forecasting framework and analyze the performance with varying training resolutions. Meteorological data covers 1,440 samples for a day and 525,600 for the full year as shown in Fig. 1. It can be observed from Fig. 1 that, the power output of PV can be divided into three different levels, namely, high, medium and low levels. The power output is at the high level in the first 100 days of the year and reached the low level of the year in the next 125 days due to the winter season. It starts increasing and reaching to the medium level as the temperature starts increasing in the coming months. Due to changes in environmental conditions and seasonal variations, the power output also varies around the year. The solar power output decreased to the lowest level in June and July. It reached the maximum level in December-January due to a surge in temperature and solar irradiation.

## *C. Inputs of Multivariate NNE Framework*

To train the predictors in NNE effectively, there is a need to identify the inputs that have effects on solar PV output. Therefore, correlation analysis is used to select the inputs. The following parameters are considered as inputs of multivariate NNE framework:

1. Historical data of PV power output (kW)
2. Solar Irradiance (W/m$^2$)
3. Wind Speed (m/s)
4. Temperature ($^{\circ}C$)
5. Humidity (%)

# III. Input Preparation with WT

## *A. Wavelet Transform*

It has been observed with the historical solar PV output profile that the time-series data contains periodic oscillations and spikes. It also contains different types of nonstationary components in the solar PV data set. It may be due to the abrupt changes in meteorological conditions and other variables. The prediction performance of a forecasting framework can be improved with the quality of training dataset. Therefore, the wavelet transform (WT) technique is applied to process the historical PV output data to enhance forecast accuracy. The WT can be divided into two main clusters namely, discrete WT (DWT) and continuous WT (CWT). The PV output can be separated into a series of constitutive components, which demonstrate more stable behavior with fewer variations and sudden spikes than the original. Therefore, Mallat's multi-resolution technique is used for pre-processing of the input [18]. The CWT can be defined as following [19]:

$$CWT_x(a,b) = \frac{1}{\sqrt{|a|}} \int_{-\infty}^{+\infty} \Psi^*(t)x(t)dt, \quad a > 0 \tag{3}$$

$$\Psi_{a,b}(t) = \frac{1}{\sqrt{|a|}} \Psi\left(t - b/a\right), \qquad a > 0, \quad -\infty < b < +\infty \tag{4}$$

where $x(t)$, $\Psi_{a,b}(t)$, $a$, $b$ and $*$ are signal for analysis, mother wavelet signal, scaling factor, translating parameter and conjugate complex parameter, respectively. High and low-frequency components of the signal provide the detailed information and non-detailed information about the signal. The DWT can be achieved by discretized scaling and translating the mother signal as follows [19]:

$$DWT_x(m,n) = 2^{-(\frac{m}{2})} \sum_{t=0}^{T-1} x(t)\Psi\left(\frac{t - n2^m}{2^m}\right) \tag{5}$$

Mallat's multi-resolution technique is used for pre-processing the input. Using high and low pass filters, the original solar PV output signal is transformed into detailed and approximation components. Then a three-level WT decomposition was applied.

# IV. Proposed NN-based Ensemble Framework

## *A. Neural Network Ensemble Framework*

The proposed framework is based on a systematic combination of FNN predictors in NNE. Neural Predictors are trained with Levenberg-Marquardt (LM) and particle swarm optimization (PSO) for better network learning. The working process of NNE is shown in Fig. 2 and explained below.

***Process 1:*** (Data preprocessing) The forecast model input variables are selected at the first stage based on correlation with PV power output. The selected input variables such as historical PV power output, wind speed, temperature, humidity and solar irradiation are used to train the NNE.

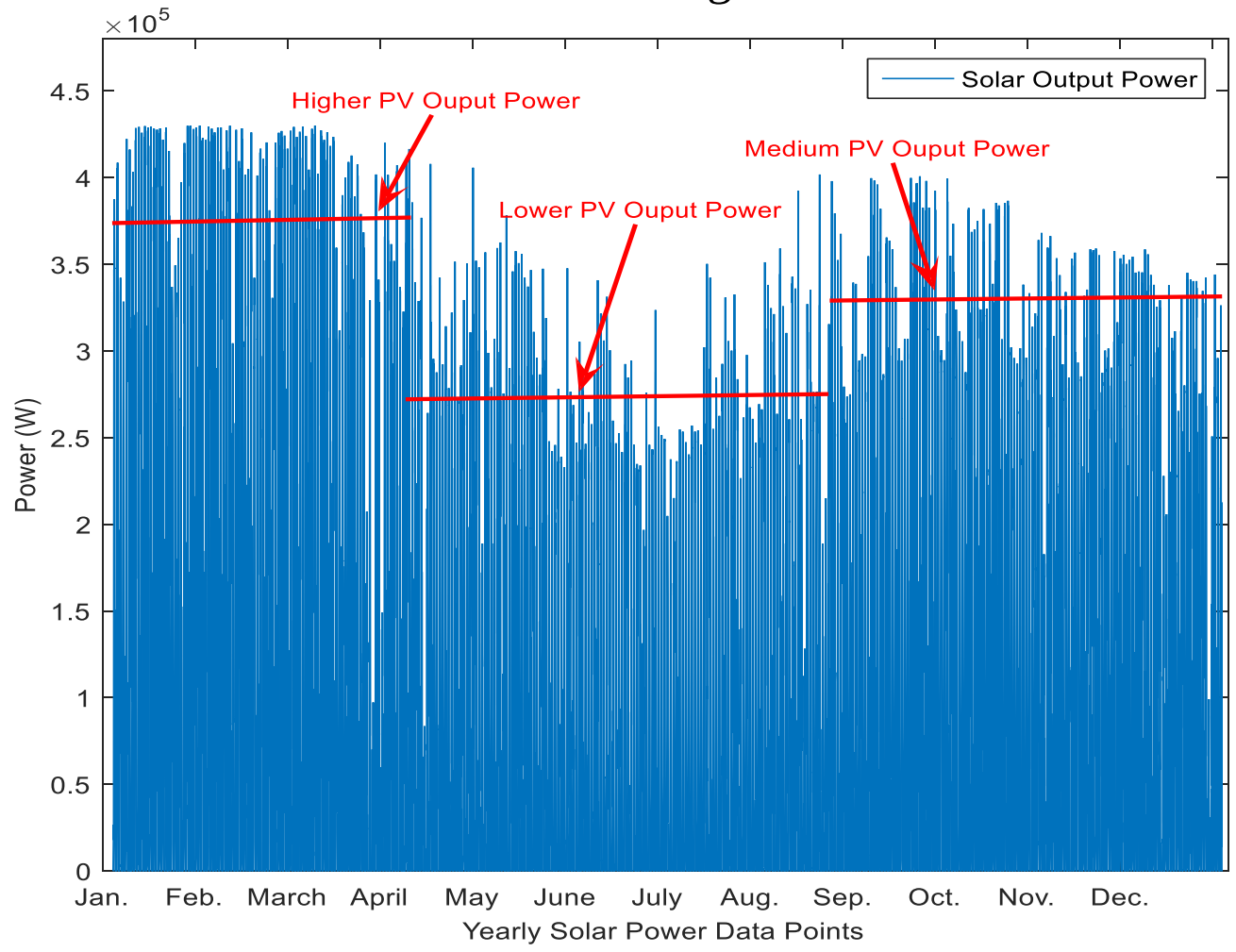

Fig. 1. Yearly (2014) Solar PV power output profile of UQ center array.

***Process 2:*** (Wavelet Decomposition) The historical PV data is pre-processed using WT to smooth the sharp changes and spikes. The WT transforms historical PV output data in approximate and detailed components as shown in Fig.3. These components are used to train predictors in the NNE framework. ***Process 3:*** (Construction of Predicators in NNE) Fig. 2 highlights the proposed NNE framework. The 5 FNN structure is initialized and each FNN structure contains 20 FNN models by completing a comprehensive performance analysis. The overall forecast performance of NNE varies with a change in the number of structures and models in each structure. In this research, the number of structures and models are selected based on a number of performance tests. In addition, another research utilizes similar structures based on load forecast application [15]. The number of NN structures in EN can be represented as in Eq. 6.

$$NN_Struc = \sum_{n=1}^{5}\{n_1, n_2, n_3, ..., n_5\} \qquad (6)$$

In each NN structure, every FNN model in the NNE framework contains the same number of network layers such as input, hidden and an output layer. The number of input neurons in each NN structure is the same as it depends on the inputs of the predicotr. However, the number of hidden layer neurons are different for each structure. In the first structure, 10 hidden layer neurons are selected. The second structure contains 5 more than the first structure, which is 15 and so on. The last or tenth structure will contain 60 number of neurons in the hidden layer of the network. The computational complexity of each NN in EN varies from one another due to different architectures. As a result, the performance of each NN and structure varies due to network architecture and the number of neurons in the hidden layer. The FNN structure is trained with LM and PSO training methods. The structures 1 and 2 are trained using the LM algorithm. The PSO is used to train the structures 3, 4 and 5. The objective is to train the FNN predictor to get the diverse forecast output.

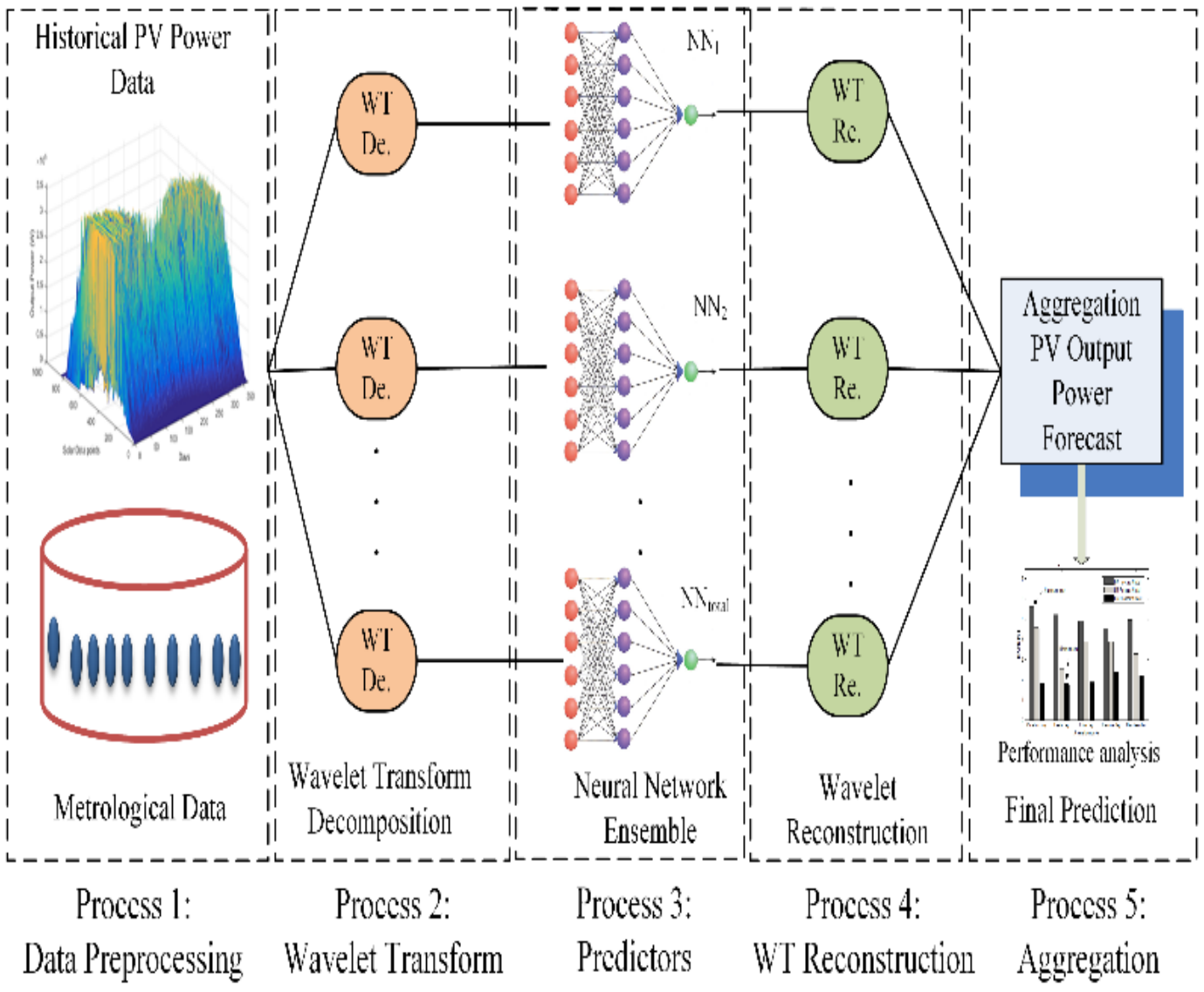

Fig. 2. Proposed NNE framework for PV output power.

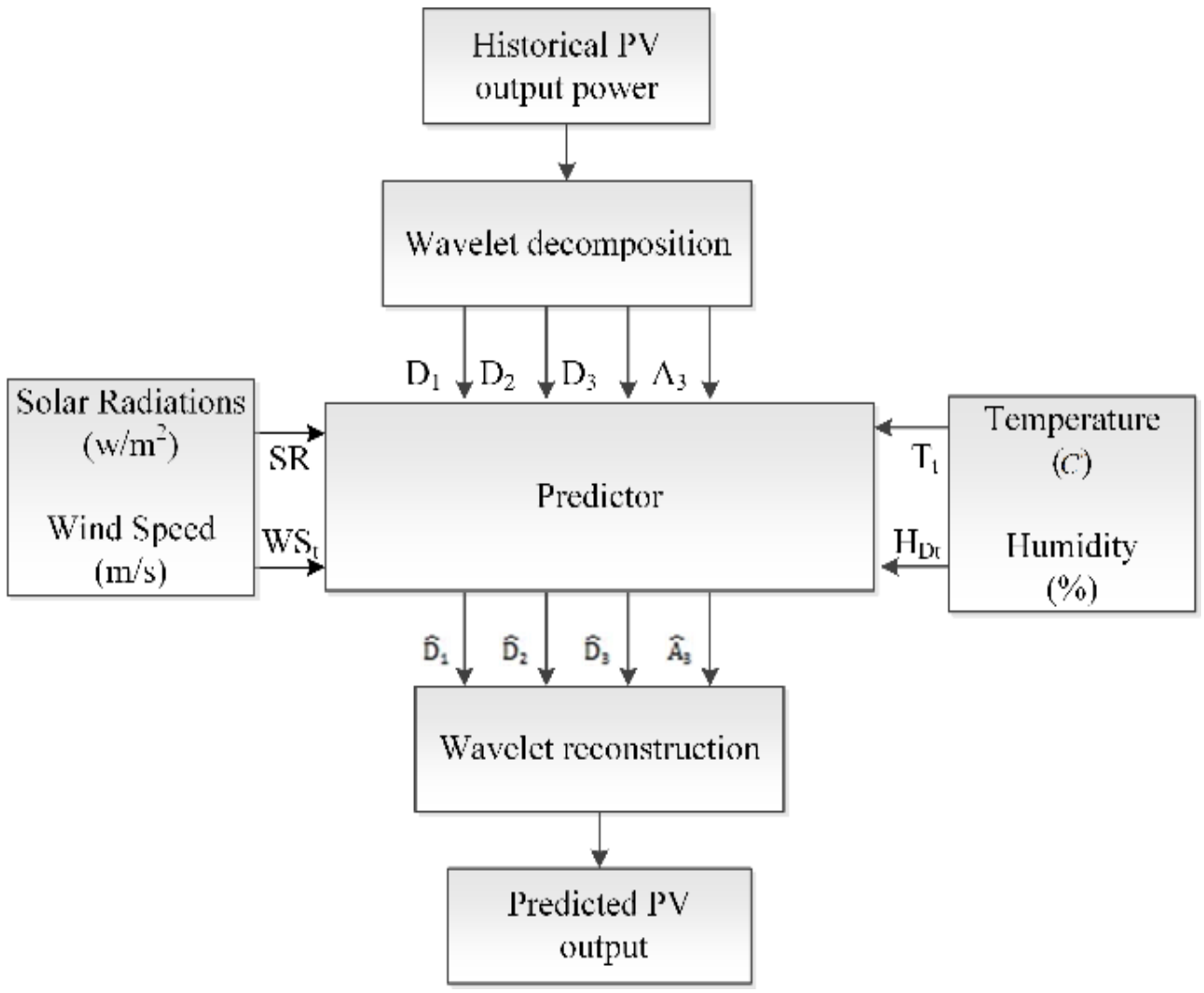

Fig. 3. Multivarite individual Predictor in NNE framework.

***Process 4:*** (Wavelet Reconstruction) The output of each FNN predictor is combined using the wavelet reconstruction process. Detailed and approximated components are combined using up and down sampling. The output of each predictor can be expressed in Eq. 7.

$$Z_{t,i} = |Output|_i^N \qquad (7)$$

where the network output is $Z_{t,i}$ and $i = \{1, 2, 3, ...., N_{total}\}$ of each predictor for all structures in the ensemble network.

***Process 5:*** Each of predictor output in NNE is combined using fusion/trim aggregation technique. The output of the aggregation technique is the final output of the NNE.

## *B. Trim Aggregation Technique of the Best NN Models*

Each predictor in the proposed NNE framework will produce the forecasting results according to its individual performance. The forecast output of NN predictors may differ from each other due to the changes in the NN architectures and network initialization parameters. In the equal weight aggregation technique, each predictor is combined in the final forecast output by averaging. The overall forecast error of the EN may increase due to the outliers of bad performing models. The forecast accuracy can be enhanced by combing the output NN predictors after removing the extreme errors. Therefore, the trimming aggregation technique is used to remove the lower and upper extreme forecasts. The trim aggregation technique was also used for grid-level load demand forecast [20], where the forecast accuracy of the model demonstrated the significance of the trimming aggregation forecast. Literature indicates that discarding the worst performing model output using the trimming parameter $\alpha$ is an effective way to remove the lower and upper extreme forecast. There is no exact method to determine the value of the trimming parameter $\alpha$. It assumed that, $\alpha \in [0,100]$. The trimming amount can be determined using network validation data by sorting them in ascending order. Then trimming number can be calculated using the trimming amount of NNE output. The output trimming can define as given Eq. 8.

$$NN_{trim} = \alpha/100 * NN_{total} * X_{\alpha} \qquad (8)$$

where $X_\alpha$ is defined as in Eq. 9.

$$X_\alpha = \left[ \frac{1}{2} NN_{trim} + 1, \frac{1}{2} NN_{trim}, \cdots, N_{total} - \frac{1}{2} NN_{trim} \right] \quad (9)$$

The mean of $\alpha$ trimmed forecast output can be calculated as given in Eq. 10.

$$Z_i^\alpha = \frac{1}{NN_{total} - NN_{trim}} \sum_{j \in I_\alpha} Z_{i,j} \quad (10)$$

where $Z_i^\alpha$ is the trimmed output with $\alpha$ the trimming parameter and $i = 1, 2, ..., D_{vald}$ . The forecast error of $\alpha$ trimmed output can be calculated as

$$MAPE^\alpha = \frac{1}{N_h} \sum_{i=1}^{N_h} \left| \frac{Z_i^\alpha - Z_i}{Z_i} \right| \quad (11)$$

where $N_h$ is forecast horizon, $Z_i$ is the traget output at the time $i$ and $z_i^\alpha$ is the trimmed output.

## C. Impact of Training Data on Forecasting

The impact of the type and length of the training data is analyzed to improve the prediction accuracy of the proposed NNE framework. Generally, the forecast accuracy of the predictor is better with longer and quality training data. The quality of training data can be affected due to peculiar meteorological conditions and anomalies in the data set. In peculiar meteorological conditions, there are large oscillations, peaks and spikes observed in solar irradiation and air temperature in peculiar meteorological conditions. Solar irradiance is highly correlated with and affects largely on PV power output. In order to analyze the quality of training data, the data is classified in different sets based on the clearance index $K_t$ . The training days are divided into three clusters, namely clear day (CD), partially cloudy day (PCD) and cloudy day (CLD) based on the clearance index $K_t$ [21]. The clearance index is the "ratio of measured global solar insolation to the calculated extra-terrestrial horizontal insolation". The extra-terrestrial horizontal insolation is the amount of solar insolation at a place without atmospheric effects on the horizontal surface. Extra-terrestrial horizontal insolation is not available in UQ solar database and it can be calculated as [22]:

$$H_0 = \frac{24 I_0}{\pi} \left[ \cos(\phi)\cos(\delta)\sin(\omega_{sr}) + \omega_{sr}\sin(\phi)\sin(\delta) \right] \quad (12)$$

$$\omega_{sr} = \cos^{-1}(-\tan(\phi)\tan(\delta)) \quad (13)$$

$$I_0 = I_{sc} \left[ 1 + 0.33 \cos\left( \frac{360N}{365} \right) \right] \quad (14)$$

where $\omega_{sr}$ is sunrise hour angle and $I_0$ is extra-terrestrial irradiance in radians and kW/m$^2$, respectively, $I_{sc}$ is solar constant, 1.367 kWh/m$^2$, $N$ is a day of the year ($N$=1 for the first day of the year and $N$=365 for the last day of the year), $\phi$ is location latitude, and $\delta$ is declination angle in degree,

$$\delta = 23.45 \sin\left[ \frac{2\pi(N-80)}{365} \right] \quad (15)$$

TABLE I: CLEARANCE INDEX RANGE FOR CLASSIFICATION

| Type of the Day | $K_t$ Range |
|---|---|
| Clear Day | $K_t > 0.45$ |
| Partially Cloudy | $0.25 < K_t < 0.45$ |
| Cloudy | $K_t < 0.25$ |

It is observed that the PV output power is highly fluctuating throughout the day due to large variations in solar irradiance and other aforementioned parameters. The days are divided into groups based on the value of the clearness index as reported in [23] and provided in Table I. Based on the clearance index, the monthly data (30 days period) of one year is divided into three distinct categories as given in Table II. It is important to mention that the clearness index cut-off values are rounded off. The calculated values may vary from original clearance index due to round-off calculation and inaccuracy of recorded data sets. The clustered monthly data based on the clearness index is applied to train the forecasting framework. The training performance of the proposed NNE framework could be changed due to training data quality. It can be observed from Table II that the clear days and partially cloudy days are higher in percentage in summer season (1-60 days) as compared to other seasons. Cloudy days are observed up to 4% during the summer season and it varies for each season and year.

The percentage of clear days is higher during the summer season. Clear days are reduced from 90% to 77% during the winter season (151-240 days) in comparison with the summer season. On another side, there is a comparatively higher number of cloudy days observed during the winter season. However, the percentage of clear, partially cloudy and cloudy days are not only depending on the season, but also on other factors such as the rainy season and an abrupt swing in meteorological conditions. As a result, sharp changes in solar irradiance can significantly affect the PV power output. Generally, the solar PV power output is higher during the summer than the winter mainly due to solar irradiance.

For better forecasting accuracy, there is a need to apply sufficient length of data for training of the forecasting framework, which at least contains some features of the forecasted time series. A higher number of mixed days are observed in winter season than in the summer. It is most likely due to forecast capability with winter training data may be better for partially cloudy and cloudy days than with the summer training data.

TABLE II: RATIO OF CD, PCD, CLD IN DATA SET

| Season | Days | Clear | Partially Cloudy | Cloudy |
|---|---|---|---|---|
| Summer | 1-30 | 81% | 16% | 3% |
| | 31-60 | 71% | 26% | 3% |
| Autumn | 61-90 | 87% | 6% | 7% |
| | 91-120 | 80% | 16% | 4% |
| | 121-150 | 77% | 12% | 11% |
| Winter | 151-180 | 60% | 26% | 14% |
| | 181-210 | 77% | 11% | 12% |
| | 211-240 | 67% | 22% | 11% |
| Spring | 241-270 | 77% | 20% | 3% |
| | 271-300 | 87% | 13% | 0% |
| | 301-330 | 87% | 10% | 3% |
| Summer | 331-360 | 90% | 6% | 4% |

## V. FORECAST RESULTS AND DISCUSSION

### A. Prediction Performance Measures

In order to analyze the performance of the proposed NNE framework, mean absolute percentage error (MAPE) and error variance $\left(\sigma^2\right)$ is calculated. The MAPE and error variance can be calculated as follows [24]:

$$MAPE(\%) = \frac{1}{FH}\sum_{t=1}^{FH}\left(\frac{PV_i^a - PV_i^f}{PV_t^p}\right)100 \tag{16}$$

$$\sigma^2 = \frac{1}{FH}\sum_{t=1}^{FH}\left(\left(\frac{\left|PV_i^a - PV_i^f\right|}{PV_t^p}\right) - \sum_{t=1}^{FH}\left(\frac{\left|PV_i^a - PV_i^f\right|}{PV_t^p}\right)\right)^2 \tag{17}$$

where *FH* is forecast horizon, $PV_i^a$ is actual PV output power, $PV_i^f$ is forecasted power and $PV_t^p$ is peak output power at a time instant *t*.

### B. Correlation between Solar Irradiance and PV Output Power

The solar PV power output has a strong positive correlation with solar irradiance. Therefore, a slight change in solar irradiance is reflected in terms of variation in solar PV power output. To observe the behavior of the PV power output with respect to solar irradiance, we analyze the pattern of PV power output of the UQ Center during the clear, partially cloudy and cloudy days. Fig. 4 shows the comparison of the solar PV power output and solar irradiance for clear, partially cloudy and cloudy days in Year 2014. The graphs depict that, the solar output follows the similar pattern as the solar irradiance curve. The solar irradiance fluctuates in clear day during the hour 13 to 14 of the day and the PV power output follows the similar pattern. In addition, fluctuations can also be observed the partially cloudy day and cloudy day as shown in Figs. 4 (b) and (c), where the PV output follows closely the solar irradiation variations. The solar irradiance is less than the 400 W/m$^2$ during most of the daytime except for hour 10 to 13. The solar irradiance in partially cloudy day reaches up to the level of 1250 W/m$^2$ during the hour 10 to 13 and the PV output power also follows the same pattern. During hour 10 to 13 of the day, solar irradiations are sharply fluctuating and the solar output curve also follows the same pattern. This indicates the strong positive correlation between the PV output and solar irradiation. It can be perceived from Fig. 4 that the solar irradiance reaches the maximum of 400 W/m$^2$ in a cloudy day which is lower than half of the clear day. In addition, an average solar PV output is relativity lower in a cloudy day. Thus, the model predication accuracy is largely affected by the higher level of fluctuations, therefore, there are higher chances that the forecast model gives high prediction error for a cloudy day in comparison with the clear and partially cloudy day.

### C. Seasonal Day-Ahead Forecast Case Study

The PV power output varies during different days and seasons of the year. Therefore, seasonal one-day ahead forecast case study is designed under different meteorological conditions. Three different days named CD, PCD and CLD are selected from Winter (W), Summer (Su), Autumn (Au), and Spring (Sp) of Year 2014. The days are classified into three different groups based on the clearness index as discussed before. The selected days may not certainly correlate to real CD, PCD and CLD due to the inaccuracy of clearness index calculation and anomalies in recorded solar and meteorological data. There is a possibility that the global solar radiations reached the threshold value of the CD, PCD or CD category during some of the hours of the selected day due to particular elevation angle and clear sky conditions. However, the average value falls under the threshold level and considered in next category. Therefore, these days may vary from an actual group of days.

Table III highlights that, for the selected season during the clear days (CD), the persistence method (M1) produces higher MAPE (15.52%, 11.51%, 14.04% and 12.20%) in comparison with the proposed NNE framework (M6) (9.50%, 7.50%, 9.18% and 6.58%) for the selected season during the clear day (CD). While for the PCD, a forecast error of the persistence model (13.23 %, 14.53 %, 12.49% and 13.58%) is also higher in comparison with the proposed forecast framework (7.99% 9.75%, 8.17% and 9.66). A similar forecast accuracy is recoded during the seasonal cloudy days. The BPNN (M2) gives more accurate results than the persistence forecast, but lower compared to the FNN+PSO (M3) model. The PSO is applied for better network training for feedforward neural network (FNN) and as a result, it will enhance the forecast accuracy. The FNN+PSO model produces a higher forecast accuracy on comparative seasonal days than the BPNN and persistence models.

The WT technique is applied to the proposed forecast framework (WT+NNE+PSO) (M6) for better forecast accuracy. The forecast results show that WT contributes to improving the prediction performance of the NNE. For example, for a summer PCD, the WT+BPNN model (M4) produces the forecast MAPE 12.16%, which is lower than the standalone BPNN network (M2) error of 12.56%. A similar kind of error reduction is observed for other forecast techniques. The WT+FNN+PSO model (M5) gives better forecast results than the BPNN, WT+BPNN and persistence models. However, it produces a higher forecast error in comparison with the proposed forecast framework (WT+NNE+PSO). The proposed forecast framework outperforms the comparative models in terms of accuracy in all seasons. The minimum forecast MAPE is observed as 6.58% on a spring CD, which is produced by the proposed NNE framework. The average forecast error for all seasonal days is calculated for the proposed and comparative models for better evaluation. The proposed forecast framework (M6) produces an average MAPE of 8.73%, which is lower than all other models. In addition, the error variance of each predictor along with the proposed forecast framework is also calculated for seasonal cloudy days. It can be observed from Table IV that the persistence model produces the highest error variance (0.4754) for a winter cloudy day. The proposed forecast model produces the PV output forecast error variance of (0.3103) for the same selected day, which is lower than those of all other comparative predictors. The FNN+PSO predictors give lower error variance than the persistence and BPNN models but higher than the proposed forecast framework. The error results indicate that the proposed forecast framework outperforms the comparative models for all selected seasonal cloudy days.Fig. 5 highlights the box forecast error plot of the proposed NNE framework along with other comparative forecast models.

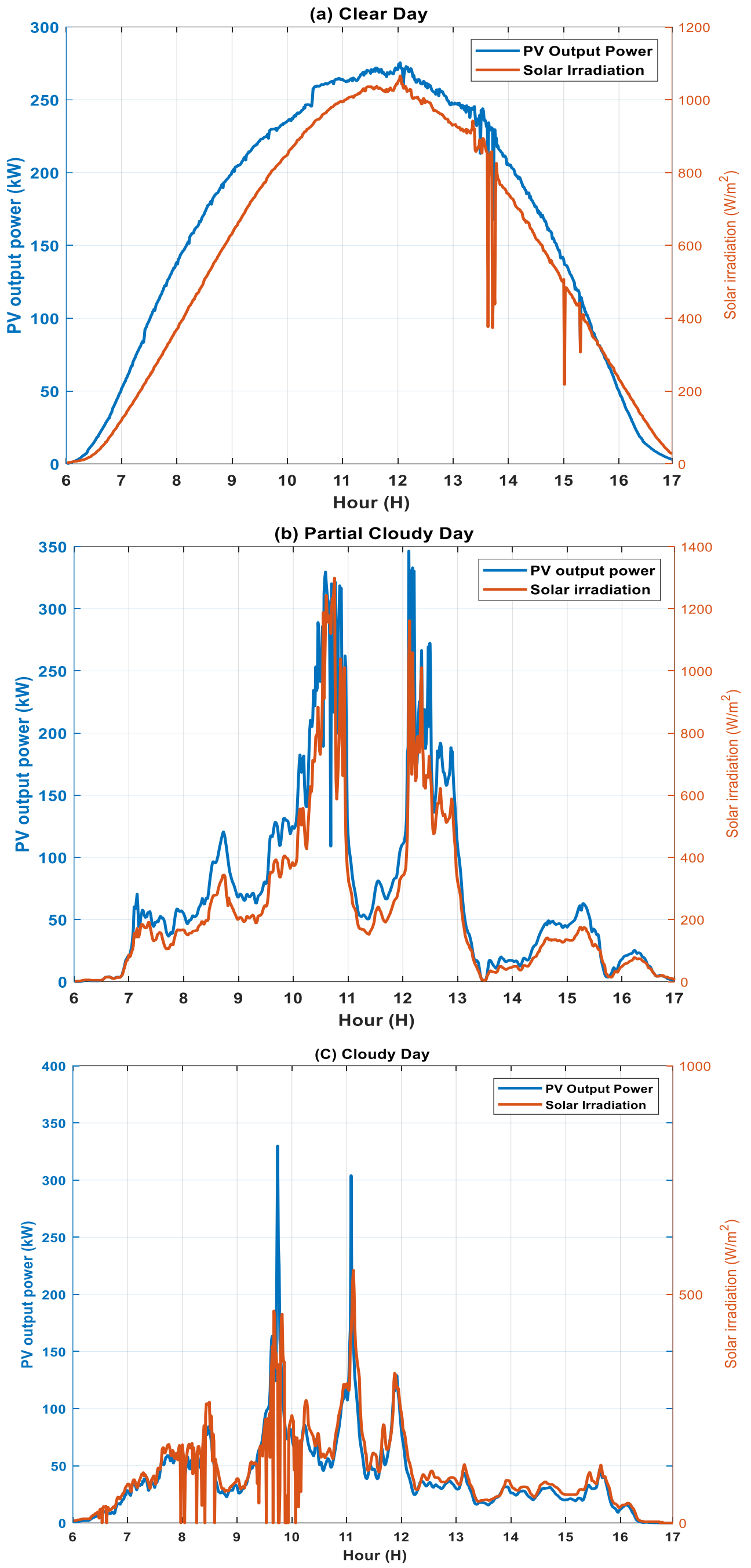

Fig. 4. The relationship between the PV output and solar radiation for (a) clear day, (b) partially clody day, and (c) cloudy day.

The median error values of the persistence technique, BPNN, and FNN+PSO are approximately 13.50%, 13.25%, and 11.45%, respectively. In addition, Fig. 5 depicts that the WT+BPNN,WT+FNN+PSO and the proposed NNE framework are producing the forecast median MAPE of 12.25%, 11.80%, and 9.75%, respectively. Forecast error comparison indicates that the wavelet transformed forecast predictors generate less forecast error in comparison with the case without WT, demonstrating the potential benefit of the WT technique in the PV output forecast application. The proposed NNE framework demonstrate less prediction error than the individual predictors and benchmark model. Forecast results indicate that the forecast performance of the proposed NNE and benchmark models vary in the seasonal day-ahead forecast. The larger size of box quartile indicates that predictors are less consistent and gives a higher error for the seasonal daily forecast. The smaller size of box quartile of the proposed NNE demonstrates that it is more accurate than other comparative models.

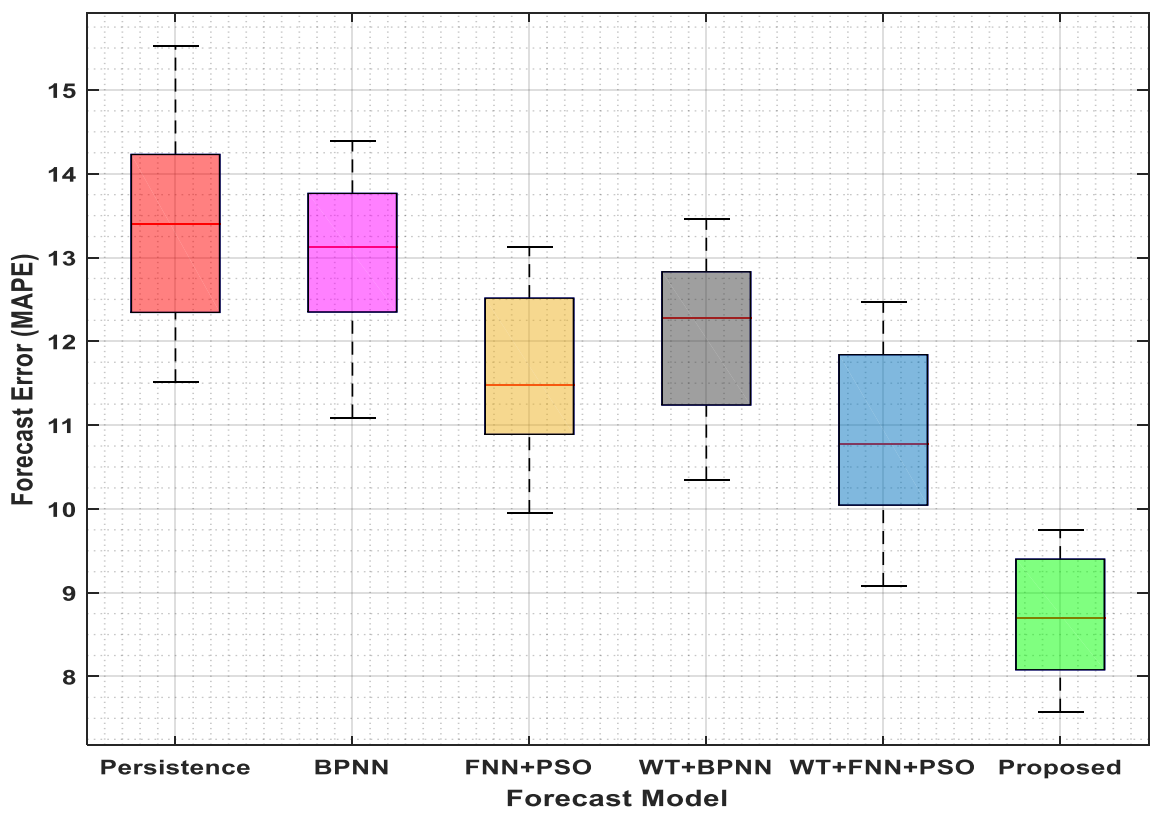

Fig. 5. Error comparison for day-ahead PV power output forecast.

Fig. 6 shows the graphical forecast comparison of all five forecast frameworks for a typical clear, partially cloudy and cloudy day. From Fig. 6, it can be observed that the persistence and WT+BPNN predictors give higher forecast errors and unable to fully capture the actual PV power output pattern during a clear day, while WT+FNN+PSO gives better forecast results than persistence and WT+BPNN predictors. However, the proposed NNE forecast model is more accurate than the comparative models and closer to the actual PV output power pattern. Similar PV output forecast pattern can be observed during the partially cloudy and cloudy day.

## D. *Impact of Training Data Resolution*

In order to analyze the impact of training data resolution, four different training data sets are prepared. These training dataset resolutions are 1, 15, 30 and 60 min. The proposed NNE framework is trained using different training datasets and performance is analyzed under different scenarios. Fig. 7 shows a regression plot for prediction performance of the proposed forecast framework with 1, 15, 30 and 60 min training datasets. The $R^2$ value between network target and output is calculated to measure the performance of each training dataset [25]. The $R^2$ value can be calculated in Eq. 18.

TABLE III: SEASONAL DAY AHEAD FORECAST ERROR COMPARISON

| Season | Day | M1 | M2 | M3 | M4 | M5 | M6 |
|---|---|---|---|---|---|---|---|
| Su | CD | 15.52 | 13.67 | 12.44 | 13.06 | 11.77 | **9.50** |
| Su | PCD | 13.23 | 12.56 | 10.86 | 12.16 | 9.96 | **7.99** |
| Au | CLD | 13.85 | 13.41 | 11.92 | 12.45 | 11.00 | **8.85** |
| Au | CD | 11.51 | 11.38 | 10.24 | 10.62 | 9.42 | **7.90** |
| Au | PCD | 14.53 | 14.39 | 13.12 | 13.46 | 12.47 | **9.75** |
| W | CLD | 12.93 | 12.84 | 11.04 | 12.39 | 10.23 | **8.33** |
| W | CD | 14.04 | 13.87 | 12.78 | 12.80 | 12.43 | **9.18** |
| W | PCD | 12.49 | 12.41 | 10.92 | 11.13 | 10.56 | **8.17** |
| Sp | CLD | 14.42 | 13.86 | 12.59 | 12.86 | 11.91 | **9.30** |
| Sp | CD | 12.20 | 12.29 | 10.92 | 11.35 | 10.13 | **6.58** |
| Sp | PCD | 13.58 | 13.41 | 12.18 | 12.06 | 11.29 | **9.66** |
| Su | CLD | 11.51 | 11.09 | 9.95 | 10.35 | 9.08 | **8.55** |
| Average | | 13.31 | 12.93 | 11.58 | 12.05 | 10.85 | **8.73** |

Su=Summer, Au=Autumn, W=Winter, Sp=Spring, M1= Persistence, M2= BPNN, M3= FNN+PSO, M4= WT+BPNN, M5= WT+FNN+PSO, M6= Proposed

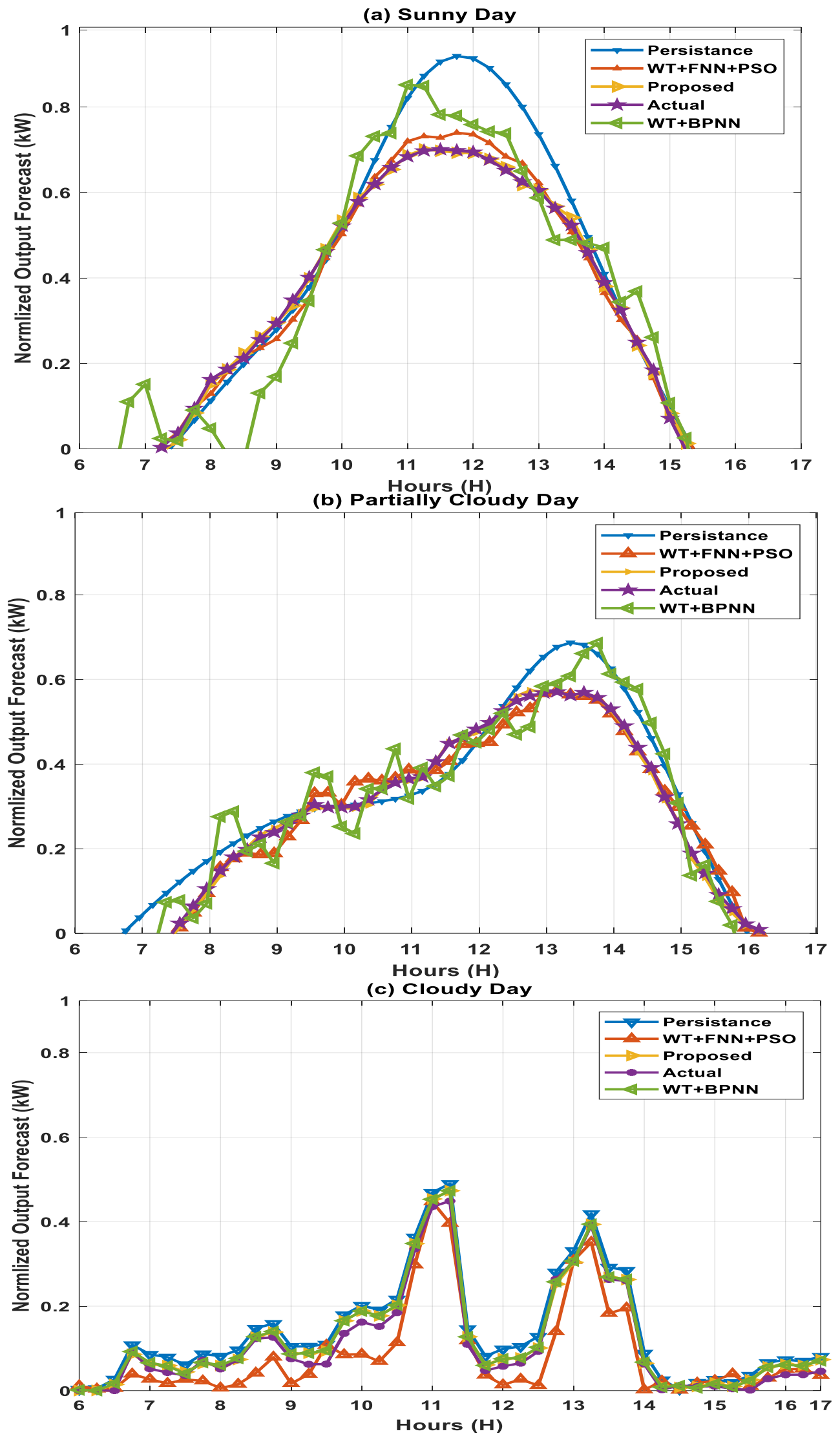

Fig. 6. One-day ahead forecast error comparison of a typical clear day, partially cloudy day, and cloudy day.

$$R^2 = 1 - \frac{\sum_{t=1}^{FH} (PV_i^a - PV_i^f)^2}{\sum_{t=1}^{FH} (PV_i^a - m_{PV_i^a})^2} \quad (18)$$

where FH is the forecast horizon, $PV_i^a$ is the actual PV power output, $PV_i^f$ is the forecasted power and $m_{PV_i^a}$ is the mean of actual PV power output. The actual PV power output is the value recorded by the data management system of UQ solar facility. The results indicate that the 1, 15, 30 and 60 min data set produces the $R^2$ value of 0.913, 0.922, 0.896 and 0.892, respectively. It can be observed that the training data with 15 min resolution produces better forecast than the 1, 30 and 60 min training datasets. This may bedue to the fact that 1 min resolution produces high variability and uncertainty than the 15 min training data set. In addition, 30 and 60 min data set cannot train the neural forecast framework properly. It is observed that predictors are not trained properly due to the uncertainty of high-resolution data or the less training patterns in low-resolution data. This unexpected behavior shows that an intermediate temporal resolution is a better option to train the network rather than the very high- or very low-resolution datasets.

TABLE IV: ERROR VARIANCE OF PREDICTORS FOR SELECTED CLOUDY DAYS

| | Summer | Autumn | Winter | Spring |
|---|---|---|---|---|
| Persistence | 0.4402 | 0.4289 | 0.4754 | 0.3024 |
| BPNN | 0.4357 | 0.4156 | 0.4435 | 0.2846 |
| FNN+PSO | 0.3574 | 0.3745 | 0.3912 | 0.2264 |
| WT+BPNN | 0.3752 | 0.3892 | 0.4123 | 0.2564 |
| WT+FNN+PSO | 0.3565 | 0.3498 | 0.3694 | 0.2145 |
| Proposed | **0.2847** | **0.2645** | **0.3103** | **0.1723** |

### E. *Impact of Training Data Length*

It is considered that the longer training data provide the better forecast output. However, the quality of training data can also affect the prediction output due to peculiar weather conditions and anomalies in the training dataset. The affected large training dataset may lead to improper training performance of the network. As discussed earlier, days can be divided into three different groups based on the clearness index. Different training datasets are created and proportion of different days as such as clear day, partially cloudy day and cloudy day are identified. Furthermore, in order to assess the performance of data quality, three different clear days are selected from each of summer, winter, and spring. To predict the PV power output of selected days, the three different datasets are applied to train the proposed NNE forecast framework with different length of training data.,where the length of datasets is 30, 60 and 90 days. Fig. 8 compares the forecast performance of the proposed framework with different training datasets. On Day-1, the forecasting framework produces the MAPE of 9.45%, 9.70% and 9.35% for 30, 60 and 90 days of training datasets, respectively. It can be observed that training error is increased when 60 days dataset is applied as compared to the 30-day data set. This may be due to the increase in partially cloudy days from 16% to 26% and the overall proportion of partially cloudy days are increased. Literature also indicates that longer training data set does not necessarily contains more training samples. This may contain repetition of similar data [26][28]. In [27], authors suggest that longer training data does not ensure better training of network despite the higher computational cost. It can be concluded that the network can be trained better with quality training data and sufficient data samples.

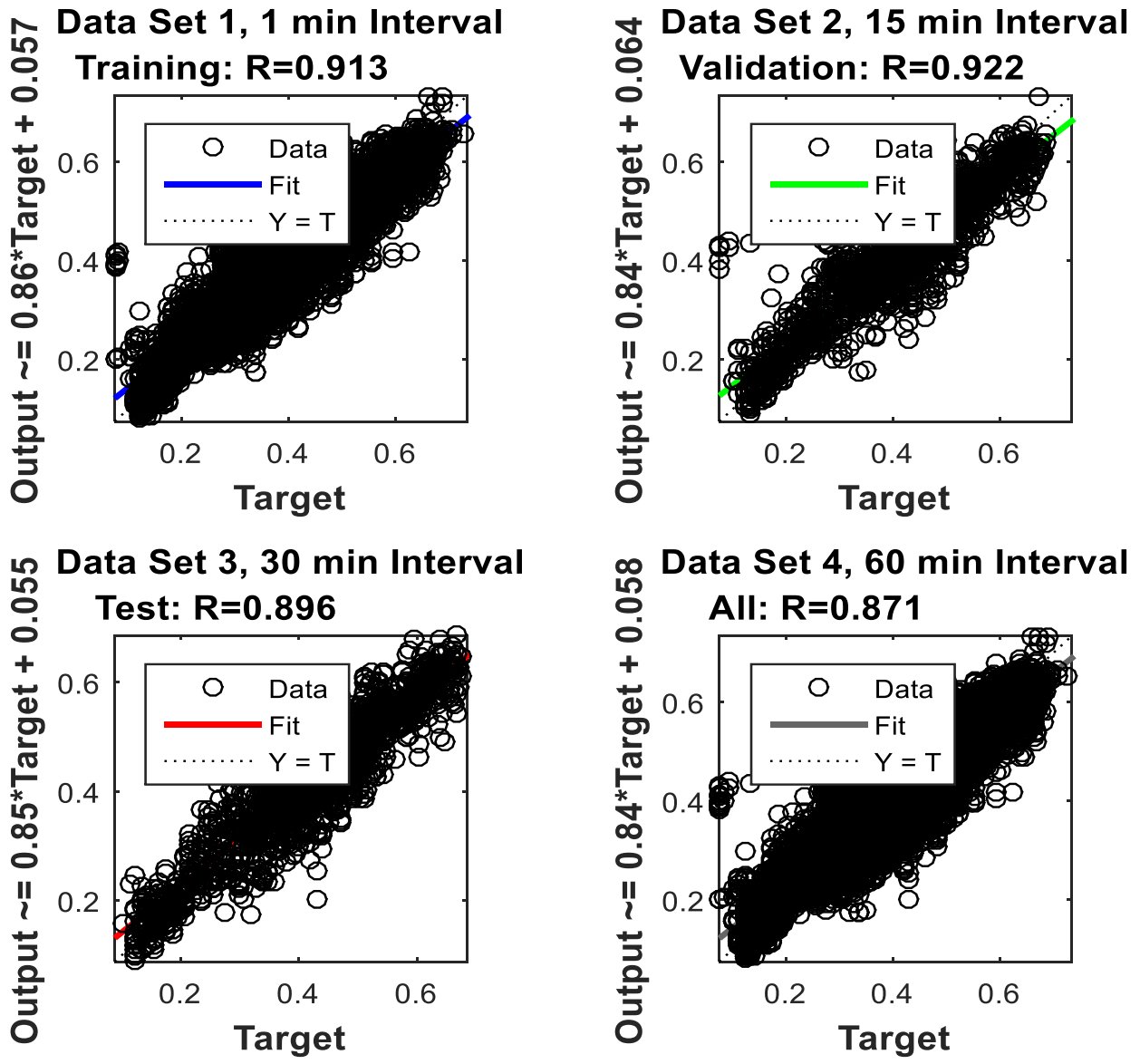

Fig. 7. Regression plot of different training data frequency.

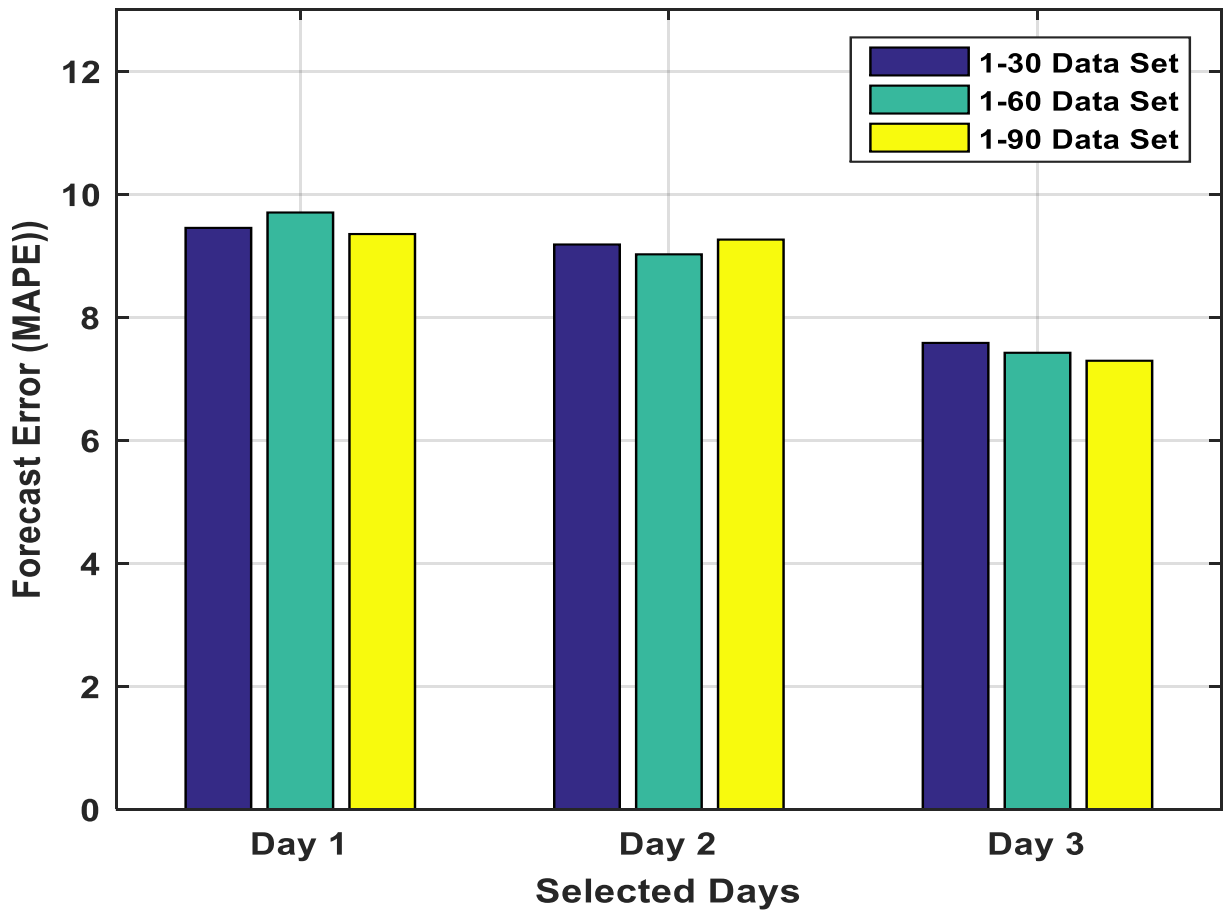

Fig. 8. Forecast performance comparison with different training data sets.

## VI. Conclusions

The performance of the proposed neural network ensemble framework is analyzed for the seasonal clear, partially cloudy and cloudy days. The predicted results of the proposed NNE forecast framework are compared with those of the comparative neural predictors and their wavelets. Results indicate that the proposed NNE framework outperforms other methods with minimum MAPE in all seasonal days. The forecast results indicate that the proposed NNE framework achieves decreases in prediction error as compared to other models. The proposed forecast framework gives less error variance for selected spring seasonal days. In addition, the forecast performance of the proposed forecast framework is also observed using different length of training dataset, and it is found that the 15-min training data produces better forecast performance than other training datasets. In order to analyze the impact of length and quality of data on training, 30, 60 and 90 days length of training data were applied. Different training and forecast performance is observed in the proposed framework with variable length of data. In terms of future work, different clustering and data filtering techniques could be applied for smoothing the input data for better training. In addition, the performance of a new training technique and ensemble could be tested. The proposed forecast framework could be applied for load demand, wind, price and other forecast applications. In addition, this forecast framework could be used to enhance the performance of building energy comfort management system, demand side management system and other energy efficiency programs.

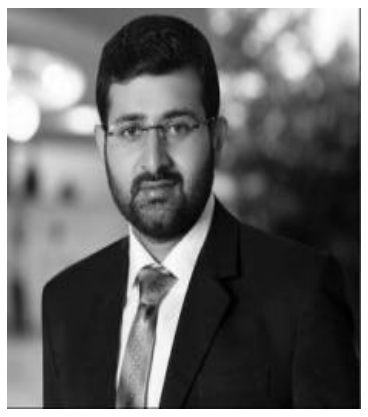

**Muhammad Qamr Raza (M’2012)** graduting with Ph.D. degree in electrical engineering from The University of Queensland, Brisbane, Australia in 2018. He received the B.Sc. degree in electrical and electronic engineering (EEE) from COMSATS IIT, Pakistan and M.Sc. by research degree in EEE from Universiti Teknologi PETRONAS (UTP), Malaysia, in in 2010 and 2014 respectivly. He is serving as associate editor of International Journal of Electrical Engineering Education (Published from UK since 1963). He also served as reviewer for several reputed journals such as IEEE Transaction on Smart Grid, Power System and IEEE conferences. His research interests includes machine learning, smart buildings, electrical load, price and PV output power forecasting.

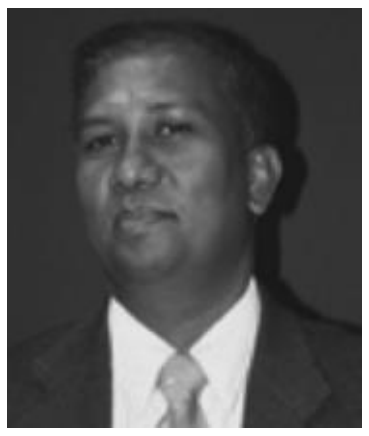

**Nadarajah Mithulananthan (SM’10)** received the Ph.D. degree in electrical and computer engineering from the University of Waterloo, Waterloo, ON, Canada, in 2002. He was an Electrical Engineer at Generation Planning Branch of Ceylon Electricity Board, Sri Lanka, and a Researcher at Chulalongkorn University, Bangkok, Thailand. He also served as an Associate Professor at the Asian Institute of Technology, Bangkok. Currently, he is with the School of Information Technology and Electrical Engineering at the University of Queensland, Brisbane, Australia. His main research interests are renewable energy integration and grid impact of distributed generation, electric-vehicle charging, and energy-storage systems.

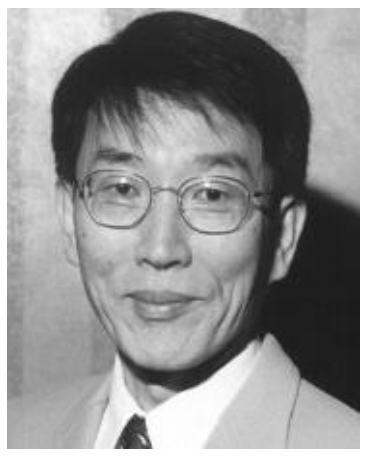

**Kwang Y. Lee** (F’01) received the B.S. degree in electrical engineering from Seoul National University, Seoul, South Korea, in 1964, the M.S. degree in electrical engineering from North Dakota State University, Fargo, ND, USA, in 1968, and the Ph.D. degree in system science from Michigan State University, East Lansing, MI, USA, in 1971. He has been in the faculties of Michigan State, Oregon State, Houston, the Pennsylvania State University, and Baylor University, where he is currently a Professor and a Chair of Electrical and Computer Engineering and the Director of Power and Energy Systems Laboratory. His interests are power systems control, operation and planning, and intelligent systems applications to power plants and power systems control.

Dr. Lee is an Editor for the IEEE TRANSACTIONS ON ENERGY CONVERSION and former Associate Editor for the IEEE TRANSACTIONS ON NEURAL NETWORK.

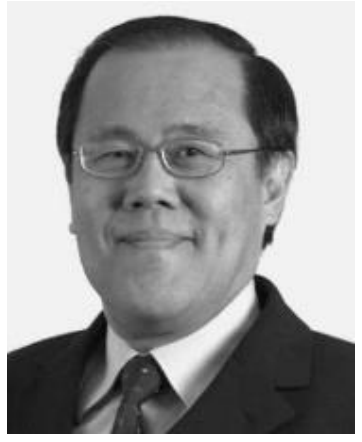

**Hoay Beng Gooi (SM)** received the B.S. degree from the National Taiwan University, Taipei, Taiwan, the M.S. degree from the University of New Brunswick, Fredericton, NB, Canada, and the Ph.D. degree from Ohio State University, Columbus, OH, USA, in 1978, 1980, and 1983, respectively, all in electrical engineering. From 1983 to 1985, he was an Assistant Professor in Lafayette College, Easton, PA, USA. From 1985 to 1991, he was a Senior Engineer in Empros (now Siemens), Minneapolis, MN, USA, where he was responsible for the design and testing coordination of domestic and international energy management system projects. In 1991, he joined the School of Electrical and Electronic Engineering, Nanyang Technological University, Singapore, as a Senior Lecturer, where he has been an Associate Professor since 1999. He was the Deputy Head of the Power Engineering Division during 2008–2014. His research interests include microgrid energy management systems dealing with storage, renewable energy sources, electricity market, and spinning reserve. He has been an Editor of the IEEE TRANSACTIONS ON POWER SYSTEMS since 2016.

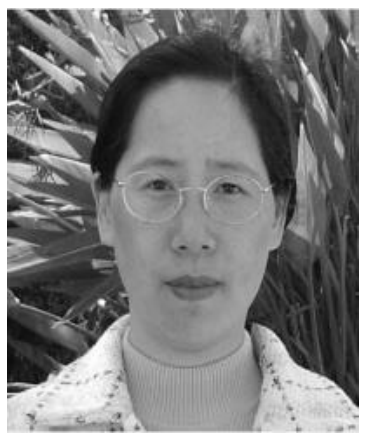

**Jiaming Li** received the B.E degree in 1986 and M.E. degree in 1989, both from the Electric Engineering Department of Nanjing University of Posts and Telecommunications, China, and PhD degree from the School of Computer Science and Engineering, the University of New South Wales, Australia in 1998.

She is currently a senior research scientist with CSIRO Data61. Her main research interests include system modelling, machine learning, data fusion, image processing, pattern recognition and multi-agent system design.